\documentclass[conference]{IEEEtran}
\usepackage{graphicx}
\usepackage{amsmath}
\usepackage{cite}
\usepackage{latexsym}
\usepackage{multirow}
\usepackage{amsfonts}

\usepackage{cite}

\ifCLASSINFOpdf

\else

\fi

\DeclareMathOperator{\real}{real}
\DeclareMathOperator{\imag}{imag}

\begin{document}
%
\title{BER Performance of Polar Coded OFDM in Multipath Fading}

\author{\IEEEauthorblockN{David R. Wasserman, Ahsen U. Ahmed, David W. Chi}
\IEEEauthorblockA{Space and Naval Warfare Systems Center Pacific\\
53560 Hull Street\\
San Diego, CA 92152\\
Email: david.wasserman@navy.mil, ahsen.ahmed@navy.mil, david.chi@navy.mil}}

\maketitle

\begin{abstract}
Orthogonal Frequency Division Multiplexing (OFDM) has gained a lot of popularity over the years. Due to its popularity, OFDM has been adopted as a standard in cellular technology and Wireless Local Area Network (WLAN) communication systems. To improve the bit error rate (BER) performance, forward error correction (FEC) codes are often utilized to protect signals against unknown interference and channel degradations. In this paper, we apply soft-decision FEC, more specifically polar codes and a convolutional code, to an OFDM system in a quasi-static multipath fading channel, and compare BER performance in various channels. We investigate the effect of interleaving bits within a polar codeword. Finally, the simulation results for each case are presented in the paper.
\end{abstract}


%
\IEEEpeerreviewmaketitle

\section{Introduction}\label{section:Introduction}
In Orthogonal Frequency Division Multiplexing (OFDM), the broadband channel is divided into several orthogonal, narrowband subchannels which are known as subcarriers. User data is modulated onto each subcarrier before transmission. The bandwidth of these subcarriers is usually smaller than the channel coherence bandwidth; therefore, OFDM systems are able to deliver high data rates with high bandwidth efficiency and robust performance in a frequency fading environment. Due to these aforementioned technical advantages, OFDM has been adopted in Wireless Local Area Network standards such as IEEE 802.11a/g/n/ac and Fourth Generation (4G) cellular technology. Although OFDM offers several advantages, it is sensitive to frequency offset and suffers from high peak average power ratio (PAPR) which is due to the nonlinearity in high power amplifiers (HPAs). \cite{Chi_Companded, Chi_Error, Chi_NBI, Ahmed_CEOFDM, Ahmed_Receiver}

In order to improve the bit error rate (BER) performance of a wireless communication system, forward error correction (FEC) codes are often implemented and utilized. FEC mitigates signal degradation by adding redundancy to transmissions. Typically an FEC encoder takes $K$ bits of user data as input, and outputs $N > K$ bits. Therefore, among the $2^N$ sequences of $N$ bits, only $2^K$ of them are valid transmissions. If the  sequence is slightly altered in transmission, the received signal is usually more similar to the transmitted sequence than to any other valid transmission, making it possible for the receiver to perfectly recover the transmitted sequence and the user data. Soft-decision FEC decoding makes use of the fact that some bits are received less ambiguously than others. Convolutional codes and turbo codes been used with OFDM \cite{Jacob_CC, Verma_CC, Ahamed_STBC}, but we have not seen any work that used polar codes with OFDM.

Polar codes are a new form of FEC introduced in 2009 by Ar{\i}kan, who proved they can achieve the capacity of any binary memoryless symmetric (BMS) channel with efficient encoding and decoding \cite{arikan}. The decoding method described in \cite{arikan} is called successive cancellation (SC). Section \ref{polar} describes polar codes in detail.

One major disadvantage of SC is that the decoder cannot correct an error made earlier in the process. In \cite{oldlist}, Tal and Vardy introduced \emph{list decoding}, also called \emph{successive cancellation list decoding (SCL)}, to mitigate this problem. They found that a list decoder's error correction performance is significantly better than SC, at the cost of higher computational complexity. They also found that performance can be further improved by using a concatenated code, with a cyclic redundancy check (CRC) encoder followed by a polar encoder. This method is called CRC-aided SCL (CA-SCL) decoding.

The paper is organized as follows. In Section \ref{section:System_Description}, the coded OFDM system under study is described in detail. Section \ref{section:Codes} describes the FEC codes and decoders that were used. Section \ref{section:Simulation_Setup} describes the simulation setup. The simulation results are shown in Section \ref{section:Simulation_Results}. Finally, Section \ref{section:Conclusion} summarizes the paper.
\section{System Description}\label{section:System_Description}
\begin{figure}[htbp]
\begin{center}
\includegraphics[width = \columnwidth]{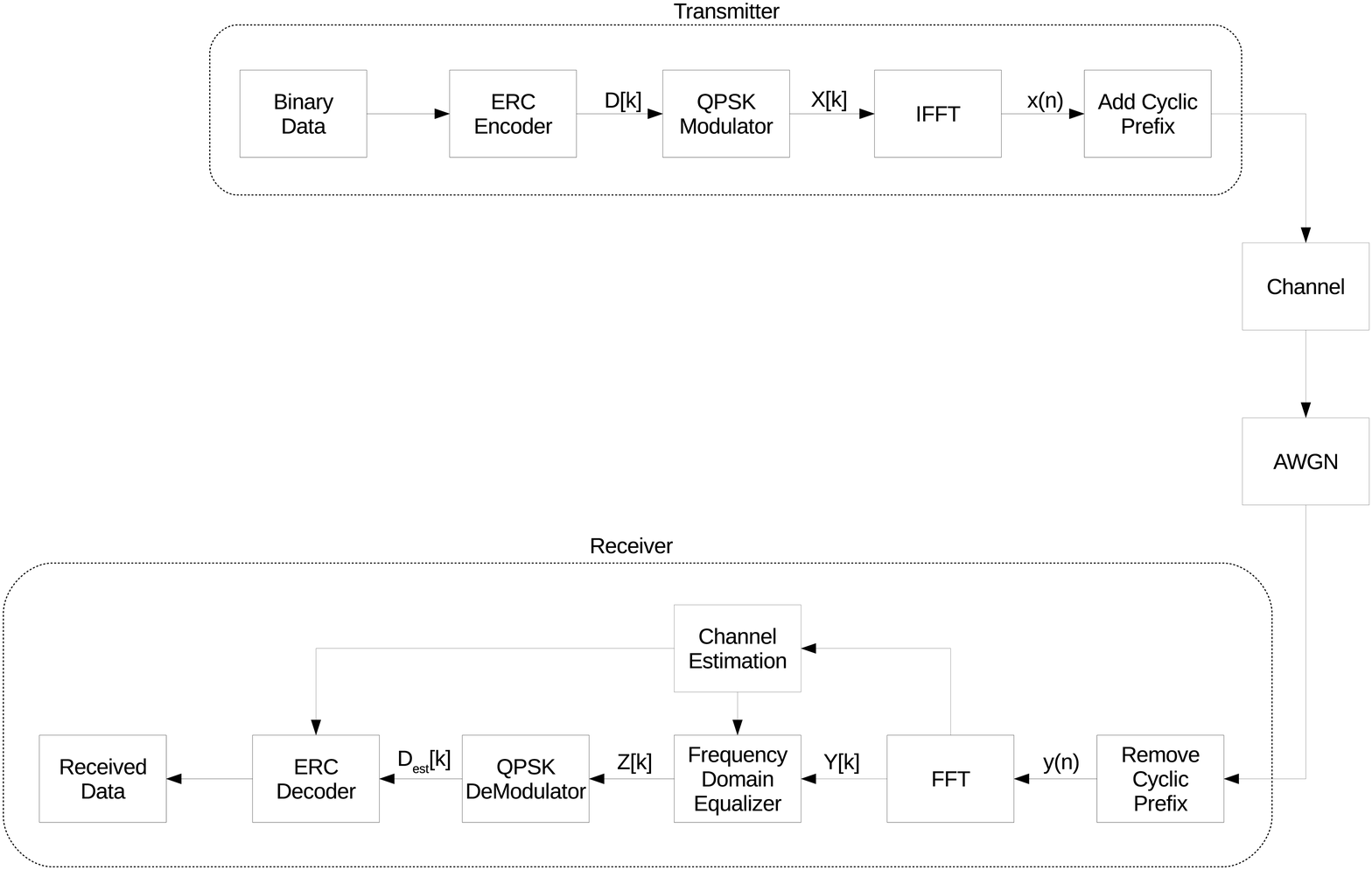}
\caption{The system block diagram of Polar Coded OFDM system in a fading channel.}
\label{fig:system_block_diagram}
\end{center}
\end{figure}

The system block diagram is shown in Fig. \ref{fig:system_block_diagram}. Each component in Fig. \ref{fig:system_block_diagram} is discussed in the following subsections.

\subsection{Transmitter}\label{subsection:Transmitter}
The transmitter consists of binary data, an FEC encoder, a Quadrature Phase Shift Keying (QPSK) modulator, an Inverse Fast Fourier Transform (IFFT), and cyclic prefix addition. The input bits are assumed to be equiprobable and statistically independent.
\subsubsection{FEC Encoder}\label{subsubsection:ERC_Encoder}
The binary input signal is first encoded with either a polar code or convolutional code. In both cases the code rate was 0.5. The polar code block size was $N = 1024$. The FEC encoder is described in Section \ref{section:Codes}.
\subsubsection{QPSK Modulator}
The output of the FEC encoder, denoted as $D[k]$, is grouped into blocks of two bits and mapped into QPSK symbols according to an alphabet, $A = \{(2m-1-\sqrt{M})+j(2n-1-\sqrt{M})\}$ where $\{m, n = 1, 2\}$ and $M = 4$ is the signal constellation size. The modulator output is called $X[k]$.
\subsubsection{IFFT}\label{subsubsection:IFFT}
At an appropriate sampling time, the signal at the output of the IFFT, denoted $x(n)$, is
\begin{align}
x(n) = \frac{1}{N_S}\sum_{k=0}^{N_S-1}X[k]e^{j\frac{2\pi nk}{N_S}}\;\;\;\;\;\;  0\leq n < N_S
\label{eqn:OFDM_signal}
\end{align}
where $N_S$ is the number of subcarriers. 

When a wireless signal is transmitted through open space, the signal is degraded by  interference, specifically intersymbol interference (ISI) which is a result of multipath propagation in the channel. To eliminate the effect of ISI, we add a cyclic prefix. The resulting signal is given by $\tilde{x}(n + N_g) = x(n)$ for $0 \leq n < N_S$ and $\tilde{x}(n) = x(n + N_s - N_g)$ for $0 \leq n < N_g$ where $N_g$ is the guard interval.

Our simulation processes two OFDM symbols on each run. The second symbol is obtained by shifting the above formulas, thus $D[N_S]$ through $D[2N_S - 1]$ determine $x(N_S)$ through $x(2N_S - 1)$ and $\tilde{x}(N_S + N_g)$ through $\tilde{x}(2(N_S + N_g) - 1)$.\subsection{Channel Model}\label{subsection:Channel_Model}
We modeled four different multipath channels, i.e. four different random processes used to generate the impulse responses, as follows:
\begin{itemize}
\item Channel A: Rayleigh fading with an exponential power delay profile, delay spread = $14T_S/N_S$, and root mean square (rms) of delay is $T_S/50$, where $T_S$ is the duration of an OFDM symbol, not including the cyclic prefix.
\item Channel B: Rayleigh fading with an uniform power delay profile, and delay spread = $14T_S/N_S$.
\item Channel C: Two Rayleigh paths, with the second path delayed by $14T_S/N_S$ and attenuated by 10 dB relative to the first.
\item Channel D: Two Rayleigh paths of equal power, with the second path delayed by $14T_S/N_S$ relative to the first.
\end{itemize}
The additive white Gaussian noise (AWGN), denoted as $w(n)$, represents the thermal noise and is modeled as an independent and identically distributed (i.i.d.) AWGN process which has zero mean and $2\sigma^2_W$ variance.
\subsection{Receiver}\label{subsection:Receiver}
Upon receiving and after removing the cyclic prefix from the signal, the received signal which is denoted as $y(n)$ is
\begin{align}
y(n) = h(n)*x(n) + w(n)
\label{eqn:received_signal_time}
\end{align}
where $h(n)$ is the channel impulse response and $*$ represents the convolution operation. The received signal is then processed by Fast Fourier Transform (FFT) and the received signal in the frequency domain, denoted as $Y[k]$, is expressed as
\begin{align}
Y[k] = H[k]X[k] + W[k]
\label{eqn:received_signal_frequency}
\end{align}
where $H[k]$ is FFT of $h(n)$.
\subsubsection{Frequency Domain Equalizer (FDE)}\label{subsubsection:FDE}
A frequency domain equalizer (FDE) is an attractive choice for OFDM equalization in frequency selective fading channels. The use of a cyclic prefix in OFDM not only mitigates any ISI but also converts a linear convolution process between the transmitted signal and the channel into a circular convolution. This circular convolution in the time domain is equivalent to simple multiplication in the frequency domain. Therefore a single tap equalizer can be used to equalize the received signal. Using the Zero Forcing algorithm, the equalized signal, $Z[k]$, is
\begin{align}
Z[k] =& \frac{Y[k]}{\hat{H}[k]} = \frac{H[k]X[k]}{\hat{H}[k]} + \frac{W[k]}{\hat{H}[k]}
\label{eqn:equalized_signal}
\end{align}
where $\hat{H}[k]$ is the channel estimate. For simplicity, we assumed perfect channel estimation, therefore $\hat{H}[k] = H[k]$, and $Z[k]$ is further simplified to
\begin{align}
Z[k] = X[k] + \frac{W[k]}{H[k]}
\end{align}
To simplify the subsequent calculations, we work with the signal $Z'[k] = Z[k]|H[k]|$ which is given by 
\begin{equation}
\label{eqn:zPrime}
Z'[k] = X[k]|H[k]| + \frac{W[k]|H[k]|}{H[k]}
\end{equation}
\subsubsection{QPSK Demodulator}
$\hat{D}[2k] = \real(Z'[k])$ and $\hat{D}[2k~+~1]= \imag(Z'[k])$ are soft estimates of $D[2k]$ and $D[2k~+~1]$, respectively.
  
For some simulation runs, an interleaver was inserted between the FEC encoder and the modulator. The corresponding deinterleaver was applied to the output of both the demodulator and the channel estimator. For each bit $D[k]$, let $H_D[k]$ be the estimate of the corresponding $H$. When no interleaver is used, we have $H[k] = H_D[2k] = H_D[2k + 1] = H_D[2(k + N_S)] = H_D[2(k + N_S) + 1]$ for $0 \leq k < N_S$.
	
\subsubsection{FEC Decoder}\label{subsubsection:ERC_Decoder}
The FEC decoder is discussed in the following section.

\section{Forward Error Correction}\label{section:Codes}
The FEC works with an information-theoretic ``channel'', denoted as $V$, that describes the probabilistic relationship between the encoder output and the decoder input, taking into account the net effect of everything in between. We call it an \emph{IT channel} to distinguish it from the wireless channel. Recall that the decoder input includes the channel estimate $H_D[k]$. Assuming perfect channel estimation, the decoder only needs the absolute value of $H_D[k]$. Thus for a particular IT channel input bit $D[k] \in \{0, 1\}$, the corresponding output is $(|H_D[k]|, \hat{D}[k]) \in \mathbf{R^+} \times \mathbf{R}$. 

The defining property of $V$ is that for any $b \in \{0, 1\}$ and $(\alpha, y) \in \mathbf{R^+} \times \mathbf{R}$, there is a conditional probability of output $(\alpha, y)$ given that the input is $b$. This conditional probability is denoted $V(\alpha, y | b)$. For optimal decoding, the decoder needs to know the \emph{likelihood ratio}, a random variable denoted as $L$, and defined as follows:
 \begin{equation}
L = \frac{V(\alpha, y | 0)}{V(\alpha, y | 1)}
\label{eq:liklihoodratio}
\end{equation}

We now show how to compute $L$ for a particular input bit $D[2k]$ from the corresponding output $(|H_D[2k]|, \hat{D}[2k])$. (The computation for $D[2k+1]$ is similar and yields the same formula.) Applying our demodulation rule to Eq. \ref{eqn:zPrime}, we get
\begin{align}
\begin{split}
\hat{D}[2k] =& \real\left(X[k]|H_D[2k]| + \frac{W[k]|H_D[2k]|}{H_D[2k]}\right) \\
=& (-1 + 2D[2k])|H_D[2k]| + \real\left(\frac{W[k]|H_D[2k]|}{H_D[2k]}\right)
\end{split}
\end{align}

The distribution of $W$ is not changed when multiplied by the unit-magnitude value $|H_D[2k]|/H_D[2k]$; it is $\mathcal{N}_{\left(0, \sigma^2_W\right)}$ where $\mathcal{N}_{(m, v)}$ is a normal distribution with mean $m$ and variance $v$. Let $P_H$ represent the probability distribution of $|H|$; then
\begin{equation}
V(\alpha, y | b) = P_H(\alpha)\mathcal{N}_{\left((-1 + 2b)\alpha, \sigma^2_W\right)}(y)
\end{equation}
and therefore 
\begin{align}
\begin{split}
\label{eq:lr}
L =& \frac{\mathcal{N}_{\left(-\alpha, \sigma^2_W\right)}(y)}{\mathcal{N}_{\left(\alpha, \sigma^2_W\right)}(y)} 
= -2y\alpha/\sigma^2_W
\end{split}
\end{align}
Note that $P_H$ cancels out, so the decoder does not need to know it. However, to predict BER, we need the probability distribution of $L$, which does depend on $P_H$. The channel response in the frequency domain, denoted as $H[k]$, is
\begin{equation}
H[k] = \sum_{n=0}^{N_S-1}h(n)e^{-j\frac{2\pi nk}{N_S}}\;\;\;\;\;\;  0\leq n < N_S
\end{equation}
The real and imaginary parts of the impulse response $h(n)$ are random variables that are independent, normal distributed, and zero-mean; see Section \ref{section:Simulation_Setup} for details. 
Therefore $|H[k]|$ has a Rayleigh distribution. The probability distribution of $L$ can then be computed as shown in Section II of \cite{boutros2013polarization}.

\begin{figure}[ht]
	\centering
		\includegraphics[width=\columnwidth]{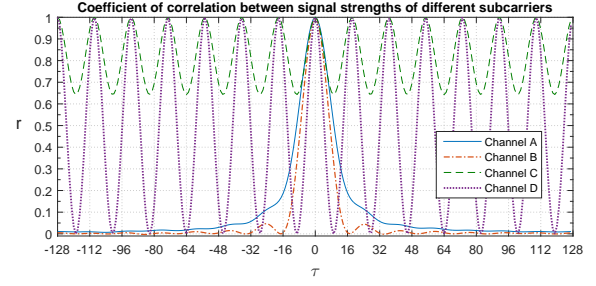}
	\caption{Coefficient of correlation between $|H[n]|$ and $|H[n + (\tau) \mathrm{mod} N_S]|$}
	\label{fig:subcarrierCorrelation}
\end{figure}

However, the OFDM IT channel is not an ergodic Rayleigh channel as described in \cite{boutros2013polarization}, because the sequence $|H_D[k]|$ ($0 \leq k < N$) is not i.i.d. In our channel models, $h(n) = 0$ for $15 \leq n < N_S$, and this results in correlations between nearby indices of the frequency response, as shown in Fig. \ref{fig:subcarrierCorrelation}.

Furthermore, each index of $H[n]$ corresponds to four indices of $H_D[k]$ with identical values.

When a block FEC code is used, its BER is determined by the joint distribution of $L(|H_D[k]|, \hat{D}[k])$ ($0 \leq k < N$), and these correlations can have a large effect on BER. Note however that the likelihood ratio, given in Eq. (\ref{eq:lr}), is not affected by these correlations: the channel estimate for the present channel bit is known, so the channel estimates for other channel bits provide no additional information about the present bit, and the noise is i.i.d. 

Similarly, the correlations do not affect the uncoded BER (although they do affect the distribution of bit errors). The uncoded BER is determined by the marginal distribution of $|H_D[k]|$, which is the same Rayleigh distribution for all $k$ and all four channel models. The probability of bit error is given by the usual formula $Q\left(\sqrt{2E_b/N_0}\right)$, where $Q(x) = \mathrm{erfc}(x/\sqrt{2})/2$. In this case $E_b = |H_D|^2$ and $N_0 = 2\sigma_W^2$. Therefore 
\begin{equation}
\label{eq:uncodedBer}
\mathrm{BER} = \int_{\alpha = 0}^\infty P_H(\alpha) Q\left(\sqrt{\alpha^2/\sigma_W^2}\right)\,d\alpha
\end{equation} 
where $P_H$ is the Rayleigh distribution. 

It will also be useful to think about the IT channel in a different way: we consider that each bit $D[k]$ passes through an AWGN IT channel $V[k]$, where the signal to noise ratio (SNR) of $V[k]$ depends on $|H_D[k]|$.

\subsection{Polar Codes}\label{polar}
Several versions of polar coding have been published. This section is intended to indicate which version is used in this work. For background and motivation of this material, see \cite{arikan}.

For any $m > 0$, we can specify a polar code of length $N = 2^m$ by choosing a subset $\mathcal{A} \subset \{1, \ldots, N\}$. If $\mathcal{A}$ has $K$ elements, we get an $(N, K)$ block code. $\mathcal{A}$ must be chosen well for good error-correction performance; we used the Tal/Vardy method from \cite{tal2013construct}. 

The polar encoder uses a row vector $\mathbf{u}$ of length $N$. Let $\mathbf{u}_\mathcal{A}$ be the subvector containing elements whose indices are in $\mathcal{A}$, and let $\mathbf{u}_{\mathcal{A}^C}$ be the subvector containing the remaining $N - K$ elements of $\mathbf{u}$. The encoder constructs $\mathbf{u}$ by filling $\mathbf{u}_\mathcal{A}$ with $K$ information bits, and setting $\mathbf{u}_{\mathcal{A}^C}$ to predetermined values, usually all 0's. These predetermined values are called \emph{frozen} bits. The encoder outputs $\mathbf{x} = \mathbf{u}\mathbf{G}_N$ where $\mathbf{G}_N$ is the $N$ by $N$ matrix $\mathbf{F}^{\otimes m}$, where $\mathbf{F} = \begin{bmatrix} 1 & 0 \\ 1 & 1 \end{bmatrix}$ and $\mathbf{F}^{\otimes m}$ is its $m$th tensor power. 

Sometimes polar coding is done using $\mathbf{F}^{\otimes m}\mathbf{\Pi}_N$ instead of $\mathbf{F}^{\otimes m}$, where $\mathbf{\Pi}_N$ is a permutation matrix called the \emph{bit-reversal operator}, defined in Section VII of \cite{arikan}. We call this \emph{bit-reversed polar coding}. In either case the encoder and decoder can be implemented with complexity $O(N \log N)$. 

\subsubsection{Systematic Polar Coding} \label{systematic}
Systematic polar coding was introduced in \cite{arikan2011systematic}. The systematic polar encoder also computes $\mathbf{x} = \mathbf{u}\mathbf{F}^{\otimes m}$ (or $\mathbf{u}\mathbf{F}^{\otimes m}\mathbf{\Pi}_N$), but the information bits are found in $\mathbf{x}$ rather than in $\mathbf{u}$. Specifically, \cite{arikan2011systematic} showed that for any vector of $K$ information bits, there is a unique $\mathbf{u}$ such that $\mathbf{u}_{\mathcal{A}^C}$ is all 0's and $\mathbf{w}_\mathcal{A}$ is the specified information bits, where $\mathbf{w} = \mathbf{u}\mathbf{F}^{\otimes m}$. The systematic SC decoder computes the estimate $\hat{\mathbf{u}}$ in the same way as the non-systematic decoder, and also computes $\hat{\mathbf{w}} = \hat{\mathbf{u}}\mathbf{F}^{\otimes m}$. Then $\hat{\mathbf{w}}_{\mathcal{A}}$ is the desired estimate of the information bits.

\subsubsection{CA-SCL}
In polar coding with CA-SCL, the polar encoder starts by computing a CRC from the information bits, and then it applies the usual polar encoding with the CRC included in the input. The list decoder produces a list of candidate codewords of the polar code, and then uses the CRC to help choose the correct one. See \cite{list} for details.

\subsubsection{Implementation}
We implemented systematic polar coding with CA-SCL. Our polar encoder and decoder software is written in MATLAB\textsuperscript{\textregistered}. The decoder is mostly based on \cite{sarkis2014increasing}. In particular, the decoder tries SC decoding first, and if the result does not pass CRC it uses list decoding. Unlike the list decoder in \cite{sarkis2014increasing}, ours computes with log-likelihood ratios (LLRs), a technique introduced in \cite{llrbasedlisttransactions}. Our decoder computes path metrics using the formula $\mu + \mathrm{ln}(1 + e^{-(1 - 2u)\lambda})$ (Eq. 11b of \cite{llrbasedlisttransactions}) instead of the piecewise-linear approximation. For SC decoding it computes LLRs using the formula $2\mathrm{atanh}(\mathrm{tanh}(x/2)\mathrm{tanh}(y/2))$ but for list decoding it uses the min-sum approximation.

\subsection{Convolution Code}\label{conv}
Convolutional codes are an FEC method introduced by Elias in 1955 \cite{elias1955coding} and are widely used in wireless communication systems to improve the BER performance. Maximum likelihood decoding of convolutional codes is efficiently achievable with the Viterbi algorithm \cite{viterbi1967error}.

\subsection{Interleaving}
Preliminary results showed that in channel A or B, the performance of the polar code could be improved by about 2 dB by interleaving within a block of $N$ bits. The interleaver was a 16-by-32 block symbol interleaver. By `symbol' we mean that each QPSK symbol is formed from a pair of bits that were already consecutive before interleaving. This interleaver puts two bits in each cell of a 16-by-32 array, filling the array in column order and outputting the results in row order.

The preliminary results were obtained using a bit-reversed polar code. When we used the same interleaver with a non-bit-reversed polar code, we found that the result was worse than without interleaving. This can be explained by observing that the non-bit-reversed polar code already includes an interleaver, and much of its effect was canceled by the block interleaver.

As explained in \cite{arikan}, polar coding is based on an operation called \emph{polarization} (or \emph{combining and splitting}) that transforms a pair of IT channels into two new IT channels. The polar decoder begins with the IT channels $V_0$ through $V_{N-1}$ that the channel bits pass through, and progresses through $\mathrm{log}_2(N)$ polarization stages. In each of these stages, polarization is applied to $N/2$ pairs, producing $N$ new IT channels. The $N$ IT channels created in the last stage are used to estimate the $N$ bits of $\mathbf{u}$. 

In the bit-reversed polar code, polarization is applied to pairs of adjacent IT channels in the first stage, and then to pairs separated by 2 in the second stage, and so on to the last stage where polarization is applied to pairs separated by $N/2$. Each IT channel formed in stage $l$ is derived from $2^l$ consecutive IT channels among $V_0$ through $V_{N-1}$. 

The non-bit-reversed polar code reverses the order, applying polarization to pairs separated by $N/2$ in the first stage and to adjacent pairs in the last stage. Each IT channel formed in stage $l$ is derived from $2^l$ IT channels among $V_0$ through $V_{N-1}$ with indices spaced by $N/2^l$.

\textbf{Question 1:} Suppose we use a length $N$ polar code with IT channels $V_0$ through $V_{N-1}$ such that some number of these IT channels are degraded relative to the others. Is it better to polarize the degraded IT channels with each other in the early stages, or in the later stages?

For example, suppose we are using a bit-reversed polar code of length 1024. In our OFDM channels, adjacent subcarriers are highly correlated, so suppose $|H[k]|$ is small for $0 \leq k \leq 7$. With no interleaving, this means $|H_D[k]|$ is small for $0 \leq k \leq 15$ and $512 \leq k \leq 527$, and the degraded IT channels are polarized together in stages 1, 2, 3, 4, and 10. With the 16-by-32 block symbol interleaver, $|H_D[k]|$ is small for $k = 16i$ and $k = 16i + 1$, $0 \leq i \leq 15$, and the degraded IT channels are polarized together in stages 1, 5, 6, 7, and 8.

Using a non-bit-reversed polar code without interleaving, the degraded IT channels are polarized together in stages 1, 7, 8, 9, and 10; with the 16-by-32 block symbol interleaver, the degraded IT channels are polarized together in stages 3, 4, 5, 6, and 10. The fact this interleaver helped in the bit-reversed case, but hurt in the non-bit-reversed case, suggests that it is better to polarize the degraded IT channels with each other in later stages. It motivates designing a deinterleaver so that IT channels from nearby subcarriers are polarized together as late as possible. For a non-bit-reversed polar code, the appropriate deinterleaver is the \emph{shuffle}, the permutation that maps $(0, 1, ..., N - 1)$ to $(0, N/2, 1, N/2 + 1, ..., N/2 - 1, N - 1)$, with the result that $H_D[4i]$ = $H_D[4i + 1]$ = $H_D[4i + 2]$ = $H_D[4i + 3]$ = $H[i]$ for $0 \leq i < N_S$. The corresponding interleaver is the inverse of this permutation, known as the \emph{reverse shuffle}. 

\section{Simulation Setup}\label{section:Simulation_Setup}
We ran 80 tests, using all combinations of the four channels models, four FEC types, and five interleavers. The FEC types were uncoded, convolutional code, polar code constructed for AWGN, and polar code constructed for ergodic Rayleigh fading. We used $(1024, 528)$ polar codes, a 16-bit CRC, and a list size of 8. The polar codes were constructed utilizing the method described in \cite{tal2013construct}, using $\mu = 512$. Some of the codes used were constructed for AWGN IT channels, and some were constructed for ergodic Rayleigh fading IT channels as defined in \cite{boutros2013polarization}. We used the CRC polynomial $x^{16} + x^{14} + x^{13} + x^{12} + x^{10} + x^{8} + x^{6} + x^{4} + x^{3} + x^{1} + 1$, obtained from Table 3 of \cite{koopman}. The convolutional code was the rate-1/2, constraint length 7 code with octal generators 133 and 171, found in many sources such as \cite{proakis}. We used the \verb#convenc# encoder and the \verb#vitdec# decoder from the MATLAB\textsuperscript{\textregistered} Communications Toolbox. We called \verb#vitdec# with $\mathrm{log}(L)$ as the input data, using opmode `cont' and decmode `unquant'. The five interleavers were
\begin{enumerate}
\item random: arbitrary permutation of length 1024, randomly chosen once for every 1024 channel bits;
\item 32-by-16 block symbol interleaver;
\item 32-by-32 block bit interleaver;
\item reverse shuffle;
\item no interleaving.
\end{enumerate}

The channel is modeled as a finite impulse response filter of length 15. For $0 \leq n \leq 14$, the in-phase and quadrature components of the impulse reponse $h(n)$ are independent Gaussian random variables with zero mean and $\sigma^2_H(n)$ variance, where $\sigma^2_H(n)$ is computed as follows:
\begin{itemize}
\item Channel A: $\sigma^2_H(n) \propto \mathrm{exp}(-n/5.12)$, normalized so $\sum^{14}_{n = 0} \sigma^2_H(n) = 1/2$.
\item Channel B: $\sigma^2_H(n) = 1/30$.
\item Channel C $\sigma^2_H(0) = 10/22$, $\sigma^2_H(n) = 0$ for $1 \leq n \leq 13$, $\sigma^2_H(14) = 1/22$.
\item Channel D $\sigma^2_H(0) = \sigma^2_H(14) = 1/4$, $\sigma^2_H(n) = 0$ for $1 \leq n \leq 13$.
\end{itemize}
A new channel impulse response, $h(n)$, is randomly generated after every two OFDM symbols (1024 channel bits), in order to sample from the long-term distribution of a slowly varying channel.

Each test began at $E_b/N_0$ = 1 dB, and the $E_b/N_0$ was increased in steps of 1 or 2 dB until a BER $< 10^{-5}$ was achieved.\footnote{From here onward, $E_b$ represents energy per information bit, not including the cyclic prefix, averaged over the probability distribution of $h$.} All polar code runs used systematic non-bit-reversed codes. Each run used a code constructed for the current $E_b/N_0$, except that for $E_b/N_0 > 17$ dB in the AWGN case, we used a code constructed for 17 dB. Finally, the number of subcarriers, $N_S$, was set to 256, and the guard interval, $N_g$, was chosen to be 15 to match the delay spread in our channel model.

\section{Simulation Results}\label{section:Simulation_Results}
For each test, we estimated the $E_b/N_0$ needed to achieve BER $10^{-5}$ by linear interpolation on a waterfall plot. The results are shown in Table \ref{tab:results}.

\begin{figure}[htbp]
	\centering
		\includegraphics[width=\columnwidth]{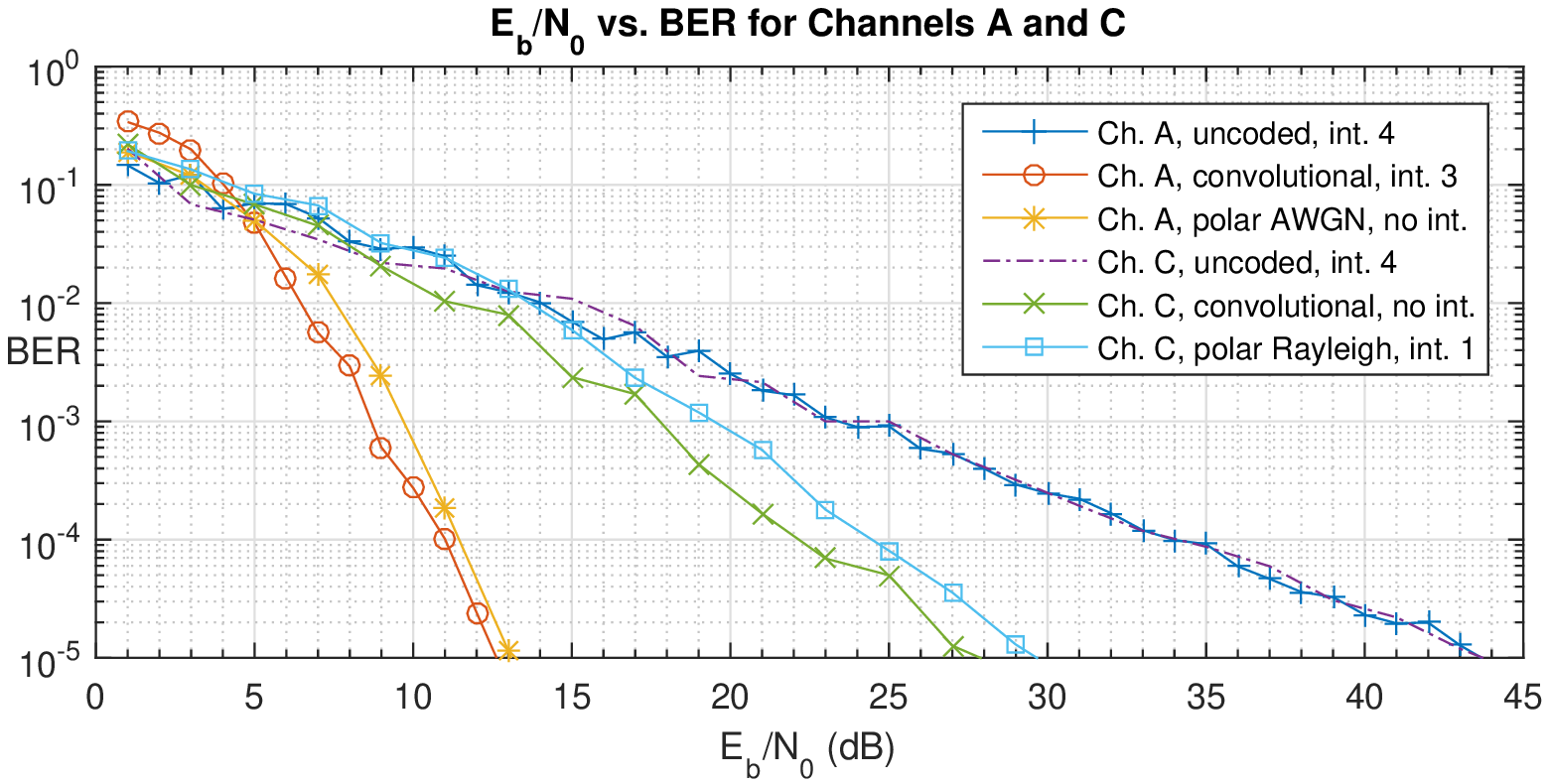}
	\caption{BER performance of coded and uncoded QPSK OFDM system in channels A and C with and without an interleaver.}
	\label{fig:chAC}
\end{figure}

\begin{figure}[htbp]
	\centering
		\includegraphics[width=\columnwidth]{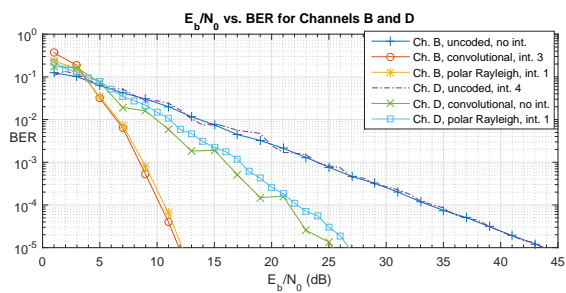}
	\caption{BER performance of coded and uncoded QPSK OFDM system in channels B and D with and without an interleaver.}
	\label{fig:chBD}
\end{figure}

\begin{table*}[t]
	\centering
	\caption{Estimated $\frac{E_b}{N_0}$ Required for BER = $10^{-5}$ for Various Channels}
		\begin{tabular}{|c|c|c|c|c|c|c|c|c|c|c|}
		\hline
                        & \multicolumn{5}{|c|}{Channel A}              & \multicolumn{5}{|c|}{Channel B }\\ \hline
                        &   1     &   2     &   3     &   4     &   5  &   1     &   2     &   3     &   4     &   5  \\ \hline
    Uncoded             &  43.9   &  43.9   &  43.8&\textbf{43.6}&  43.7&  43.9   &  44.3   &  44.0   &  43.9   & \textbf{43.8}\\ \hline
    Convolutional       &  12.7   &  13.1&\textbf{12.4}&  29.1   &  23.7&  12.3   &  12.6&\textbf{11.9}&28.2   & 22.7\\ \hline
    Polar for AWGN      &  14.4   &  18.5   &  19.5   &  16.1&\textbf{13.1}&13.4   &  17.5   &  18.3   &  15.3   & 12.3\\ \hline
    Polar for Rayleigh  &  14.7   &  18.3   &  18.3   &  15.4   &  14.1&\textbf{12.2}&17.4   &  17.6   &  14.4   & 13.1\\ \hline
                        & \multicolumn{5}{|c|}{Channel C}              & \multicolumn{5}{|c|}{Channel D }\\ \hline
                        &   1     &   2     &   3     &   4     &   5  &   1     &   2     &   3     &   4     &   5  \\ \hline
    Uncoded             &  43.9   &  43.8   &  44.0&\textbf{43.7}&44.0&  43.9   &   43.9  &  43.8&\textbf{43.6}&43.7\\ \hline
    Convolutional       &  28.9   &  29.0   &  29.1   &  28.6&\textbf{27.9}&27.0   &  26.4   &  27.4   &  26.4&\textbf{25.4}\\ \hline
    Polar for AWGN      &  29.8   &  30.2   & 30.9    &   30.6  &  29.9&  26.8   &  28.2   &  28.1   &  28.1   &  27.7  \\ \hline
    Polar for Rayleigh&\textbf{29.7}&30.6   & 30.7    &  30.8   &  29.9&\textbf{26.7}&28.5   &  28.0   &  28.6   &  27.6 \\ \hline
    \end{tabular}
	\label{tab:results}
\end{table*}
A bold number in the Table \ref{tab:results} indicates that value is the best uncoded result, best convolutional result, or best polar result for that channel. The BER curves corresponding to the bold numbers are shown in Figs. \ref{fig:chAC} and \ref{fig:chBD}. 

\subsection{Discussion}
As expected, all uncoded tests produced similar results. Based on Eq. (\ref{eq:uncodedBer}), the BER $10^{-5}$ can be achieved at SNR 44.0 dB. The small deviations from this value are due to sampling error.

In Channels A and B with convolutional codes, it is not surprising that the block bit interleaver is the best and the reverse shuffle is the worst. With this encoder, each input bit affects 10 channel bits, which are all included in a group of 14 consecutive channel bits. Errors can easily occur if all of these bits are transmitted on low-energy subcarriers. With the block bit interleaver, any two of these bits are at least 16 subcarriers apart. With the reverse shuffle, all 14 bits are on four consecutive subcarriers. 

With polar codes, the data refutes our hypothesis that it is best to polarize the degraded IT channels with each other as late as possible. It may be helpful to investigate Question 1 in a simpler scenario, in which the degraded IT channels are all equal, and the non-degraded IT channels are all equal. An interleaver that is optimal in this case may also work well in the OFDM channel models we studied.

We also observe that neither the AWGN polar code nor the Rayleigh polar code consistently outperforms the other. It may be possible to design a polar code specifically for one of these channel models.

For channel D, and even more so for channel C, we see that it makes little difference which interleaver or code we use. This is because in these channels the total energy of all subcarriers frequently falls far below average, making reliable decoding impossible. We will call this situation \emph{deep fade}. When the average total energy is high, deep fade causes most of the bit errors. 

Strategies to handle deep fade depend on latency requirements, how long the fade lasts, and whether the transmitter is able to adapt to it (for example, by adjusting the data rate.) Without such adaptation, we need to add diversity. For example, we could increase time diversity with an interleaver several times longer than the fade duration. If the interleaver is long enough, the channel approximates an ergodic Rayleigh fading IT channel, at the cost of increased latency. We have simulated an ergodic Rayleigh fading IT channel, and found that the polar code and the convolutional code achieve BER $10^{-5}$ at $E_b/N_0$ 5.1 and 9.1 dB respectively. Alternatively, \cite{7541294} showed that polarization occurs in a large class of IT channels with memory, and therefore polar codes can achieve the IT channel capacity. However, the IT channel capacity is achieved in the limit as $N \rightarrow \infty$. Large block sizes are another way to increase time diversity at the cost of increased latency.

\section{Conclusion}\label{section:Conclusion}
When used with OFDM in a variety of quasi-static multipath fading channels, the BER performance of the polar codes we have studied showed no coding gain over a simple convolutional code. However, it may be possible to improve the performance of polar codes in these channels with better code construction and interleaving methods.

\section*{Acknowledgment}
This work was funded by the Naval Innovative Science and Engineering (NISE) Program at Space and Naval Warfare Systems Center Pacific.
%



%
\bibliographystyle{styles/IEEEtran}

\end{document}